\documentclass[10pt, a4paper]{article}
\usepackage[utf8]{inputenc}
\usepackage{graphicx}	% Including figure files
\usepackage{amsmath}	% Advanced maths commands
\usepackage{todonotes}
\usepackage{amssymb}	% Extra maths symbols
\usepackage{amsfonts}
\usepackage{latexsym}
\usepackage{hyperref}	% Permits to insert hypertext links
\usepackage{mathtools}
\usepackage{rotating}
\usepackage{epsfig}
\usepackage{tabularx}
\usepackage{natbib}
\usepackage{xcolor}
\usepackage{lineno} % for line numbering
\usepackage{aas_macros}
\usepackage[
left=1.5cm,
right=1.5cm,
top=2.00cm,
bottom=2.00cm
]{geometry}
\usepackage{comment}
\usepackage{authblk}
\usepackage{titlesec}
\usepackage[labelfont=bf,figurename=Fig.,font=small,labelsep=period]{caption}
\usepackage{abstract}
\usepackage{bm}
\usepackage[T1]{fontenc}
\usepackage[english]{babel}
\usepackage{svg}
\usepackage{tikz}

\def\n{\mathtt n}
\def\k{\mathtt{k}}

\def\K{\mathcal K}

\def\K{\mathcal{K}}

\title{\huge{\textbf{The dynamical structure of the Earth co-orbital region and implications for the near-Earth asteroid population}}}

\author[1,2]{M. Fenucci\thanks{Email: \texttt{marco.fenucci@ext.esa.int}}}

\author[3]{\'{O}. Rodr\'{i}guez}

\author[4]{M. Saillenfest}

\author[1]{L. Faggioli}

\affil[1]{Planetary Defence Office / NEO Coordination Centre, ESA ESRIN, Largo Galileo Galilei 1, Frascati, 00044, RM, Italy}
\affil[2]{Deimos Italia s.r.l., Via Alcide De Gasperi 24, San Pietro Mosezzo, 28060, NO, Italy}
\affil[3]{Departament de Matemàtiques, Universitat Politècnica de Catalunya, Av. Diagonal 647, 08028 Barcelona, Spain}
\affil[4]{LTE, Observatoire de Paris, Université PSL, Sorbonne Université, Université de Lille, LNE, CNRS, 61 Avenue de l’Observatoire,
75014 Paris, France}

\date{\today}

\begin{document}
% Activate line numbering

\maketitle

% ABSTRACT, MATH SUBJECT & KEYWORDS 
\begin{abstract}

We investigate the structure of the Earth co-orbital region at low eccentricity and low inclination using a semi-analytical model of the 1:1 mean-motion resonance. The dynamics of asteroids in co-orbital motion with the Earth is described through a resonant semi-secular Hamiltonian, allowing the classification of orbits into circulation, Trojan, horseshoe, and quasi-satellite states. By systematically exploring the phase space in the space of the orbital elements, we compute the fraction of each type of motion and quantify how different co-orbital states fill the Earth co-orbital region. We find that horseshoe orbits dominate the phase space, occupying more than half of the volume, followed by Trojan and circulating orbits, while quasi-satellites represent only a small fraction. The distribution of co-orbital states exhibits strong inhomogeneities, particularly as a function of the argument of perihelion, with clear concentration regions of horseshoe orbits associated with node-crossing geometries. 
We also study the short-term stability of this portion of phase space by means of the MEGNO indicator, and how the level of chaos differs between different co-orbital states.
Finally, we discuss the implications of these results for the expected population of Earth co-orbitals and for planetary defence, showing that a large fraction of co-orbitals remains undiscovered.

\vskip0.1truecm
\noindent
\textbf{Keywords:} Earth co-orbitals - mean motion resonances - near-Earth asteroids 
\end{abstract}

\section{Introduction}
% Co-orbital motion in general
A small object with an orbital period similar to that of a planet is often called a co-orbital, since it shares a close orbital path to that of the planet itself. The proof of the existence of co-orbital motion dates back to 1772, when Lagrange demonstrated the existence of the stable equilibrium points in a triangular configuration in the circular restricted three-body problem. Today, these points are commonly known as L4 and L5 equilibrium points. 
% First Jupiter Trojan
The existence of asteroids near the stable Lagrange points was confirmed in 1906, when the first Jupiter Trojan (588) Achilles was discovered by \citet{wolf_1906}. For a historic overview on the discovery of Jupiter Trojans, the interested reader can refer to \citet{connors_2024}. Since 1906, many Jupiter Trojans were found, and nowadays we know about 15~000 of them. Studying their composition and orbital history may give fundamental answers on the formation and early evolution of the Solar System, and the NASA Lucy mission \citep{levison-etal_2025} is currently flying and scheduled to visit eight Jupiter Trojans.      

In the first decades of 1900 also horseshoe orbits \citep{brown_1911} and quasi-satellite orbits \citep{jackson_1913} were theoretically predicted using the circular restricted three-body problem. With the discovery of a large amount of asteroids in the last three decades, the study of the co-orbital motion became again a trend in the astronomical community. 
Co-orbitals were found to have complex dynamics, undergoing switches between different types of motion \citep{namouni_1999, sidorenko_2018}, and models to investigate the co-orbital dynamics at high eccentricity and inclinations were developed \citep{namouni-etal_1999, nesvorny-etal_2002}, as well as retrograde co-orbital motions \citep{sidorenko_2020}. The stability of co-orbitals over long time-scale became also a topic of interest, and several real cases were investigated using either averaged models or full $N$-body numerical simulations \citep[see for example][]{mikkola-innanen_1997, christou_2000, cuk-etal_2012, delafuente-delafuente_2016b, delafuente-delafuente_2016, pousse-etal_2017, christou_2021, santana-ros-etal_2022, pousse-alessi_2022, diruzza-etal_2023}.
Nowadays, except the large population of Jupiter Trojans, about 180 asteroids are known to be in a co-orbital motion with a planet \citep{pan-gallardo_2025}.

% Earth co-orbitals
In particular, asteroids in a co-orbital motion with the Earth are of special interest for several reasons. These objects are naturally categorized as near-Earth asteroids (NEAs), because their perihelion distance is naturally smaller than 1.3 au, and may be therefore at risk of impacting Earth. In fact, on one hand the resonance protects the Earth from impacts to occur \citep{milani-baccili_1998}, while on the other hand gravitational perturbations cause transitions between different co-orbital state, which may lead the orbit of the asteroid to cross that of the Earth \citep{fenucci-etal_2022}. 
%  Known and expected population
The first discovery of an Earth co-orbital, (3753) Cruithne, dates back to 1986, and since then many more have been found. However, NEA population models \citep{granvik-etal_2018, nesvorny-etal_2024} predict the existence of hundreds of Earth co-orbitals, thus the known population results to be largely incomplete. This is because, generally, Earth co-orbitals have a long synodic period, spending much of their time at low solar elongations, making the discovery either challenging or even impossible by ground-based telescopes \citep{morais-morbidelli_2002}. Quasi-satellites are the only exception, and survey simulations showed that the population should be 80\% complete up to absolute magnitude of 24 \citep{fenucci-etal_2026}. The limitation of ground based telescopes may be mitigated by the upcoming NEO Surveyor \citep{mainzer-etal_2023} mission by NASA and the NEOMIR \citep{conversi-etal_2023} space telescope by the European Space Agency (ESA), which will be able to observe at low solar elongations, being based on infrared observations. 
In addition to the impact threat, Earth co-orbitals generally need a low $\Delta V$ budget to be reached by a spacecraft, and they are therefore good candidates for asteroid exploration or exploitation missions. In recent year, the Earth quasi-satellite (469219) Kamo`oalewa \citep{delafuente-delafuente_2016} was selected as the target for the Tianwen-2 sample-return mission \citep{zhang-etal_2021,  zhang-etal_2025natas} of the China National Space Administration (CNSA). The spacecraft was launched on 29 May 2025, and it is expected to return a sample of Kamo`oalewa by 2027. 

In this work, we focus on the Earth co-orbital region at low eccentricity and low inclination with a two-fold objective. First, we aim at characterizing the geometric structure of this portion of phase space, quantifying how frequently different types of co-orbital motion occur and how they fill the available volume. Second, and more broadly, we use these results to estimate the expected population of undiscovered Earth co-orbitals and to discuss their implications for planetary defence. The Earth impact risk is also interpreted by means of full numerical integrations aimed at studying the stability of different classes of Earth co-orbitals.
The paper is structured as follows. In Sec.~\ref{s:methods} we introduce a semi-analytical model for the 1:1 mean-motion resonance with the Earth, and then we compute the volume of phase space occupied by each type of motion (Sec.~\ref{s:results}). Stability properties of different co-orbital types are studied in Sec.~\ref{s:megno} using numerical integrations.
The phase space fractions derived in Sec.~\ref{s:methods}, combined with modern debiased NEA population models, allow us to estimate the expected number of Earth co-orbitals of each type, while stability properties give information about the Earth impact risk. These implications are discussed in Sec.~\ref{s:discussion}. We finally provide conclusions in Sec.~\ref{s:conclusions}.

\section{Exploring the Earth co-orbital region}
\label{s:methods}
\subsection{The resonant semi-secular Hamiltonian}
\label{ss:secHam}
We model the motion of an asteroid in a mean-motion resonance with a planet by averaging over the fast angles, using a semi-secular Hamiltonian \citep{milani-baccili_1998}. Following \citet{fenucci-etal_2022}, we model the Solar System with the planets from Mercury to Neptune placed on circular and co-planar orbits centered at the Sun. 
Let us also denote with $a, e, I, \omega,\Omega,\ell$ the Keplerian orbital elements of the asteroid which are, respectively, the semi-major axis, the eccentricity, the inclination, the argument of perihelion, the longitude of the ascending node, and the mean anomaly. 
Let us assume that the asteroid is close to the $h:h_p$ mean-motion resonance with the planet $p \in \{1, \dots, 8\}$, where $h,h_p$ are co-prime integers. The angle $\sigma$ associated to the resonance is of the form
\begin{equation}
    \sigma = h \lambda - h_p \lambda_p - (h-h_p)\varpi,   
    \label{eq:resonantAngle}
\end{equation}
where $\lambda=\ell+\omega+\Omega, \, \lambda_p=\ell_p+\omega_p+\Omega_p$ are the mean longitudes of the asteroid and the planet, and $\varpi = \omega+\Omega$ is the longitude of the perihelion of the asteroid.
The resonant semi-secular Hamiltonian $\mathcal{K}$ for the motion of the asteroid is then given by 
\begin{equation}
    \mathcal{K} = \mathcal{K}_0 + \epsilon(\mathcal{K}_{\text{sec}} + \mathcal{K}_{\text{res}}), \quad \epsilon = \mu_5,
    \label{eq:semisecHamiltonian}
\end{equation}
where
\begin{equation}
    \begin{split}
        \mathcal{K}_0 & = -\frac{\k^2}{2a} - \n_p \frac{h_p}{h}\k \sqrt{a}, \\
        \mathcal{K}_{\text{sec}} & = 
        -\frac{\k^2}{\mu_5} \sum_{\substack{j=1 \\ j \ne p}}^8 \frac{\mu_j}{(2\pi)^2}\int_0^{2\pi}\int_0^{2\pi}\frac{1}{|\mathbf{r}-\mathbf{r}_j|}\text{d}\ell \text{d}\ell_j, \\
        \mathcal{K}_{\text{res}} & = -\frac{\k^2}{\mu_5} \frac{\mu_p}{2\pi}\int_0^{2\pi} \bigg( \frac{1}{|\mathbf{r}-\mathbf{r}_p|} - \frac{\mathbf{r}\cdot\mathbf{r}_p}{|\mathbf{r}_p|^3}\bigg)\text{d}\gamma.
    \end{split}
    \label{eq:semisecularHamiltonianTerms}
\end{equation}
In Eq.~\eqref{eq:semisecularHamiltonianTerms}, $\k = \sqrt{\mathcal{G}m_0}$ is the Gauss constant, $\mathcal{G}$ is the universal gravitational constant, $m_0$ is the mass of the Sun, and $\mu_j=m_j/m_0, ~ j=1,\dots,8$ where $m_j, ~ j=1,\dots,8$ are the masses of the planets. In addition, $\mathbf{r}$ and $\mathbf{r}_j, j=1,\dots,N$ are the heliocentric positions of the asteroid and the planets, respectively, $\n_p$ is the mean motion of the planet $p$, and $\gamma$ is a fast angle conjugated to the angle $\sigma$. A \texttt{fortran} program to compute the averaged resonant Hamiltonian is available on GitHub\footnote{\url{https://github.com/Fenu24/Resonant-Hamiltonian}}, and it uses the \texttt{quadpack} library to compute integrals \citep{piessens-etal_1983}.

\citet{fenucci-etal_2022} showed that NEAs in a mean-motion resonance with a planet generally undergo an adiabatic evolution, where $a$ and $\sigma$ evolve on a semi-secular timescale, while $e,I,\omega$ and $\Omega$ evolve on a secular timescale. On a short timescale, therefore, we can assume that the elements $e,I,\omega,\Omega$ are fixed, and the motion is described by the level curves of $\K$ in the $(a,\sigma)$ plane. This assumption was also successfully used by \citet{pan-gallardo_2025} to build a catalog of co-orbital asteroids in the Solar System. 

\subsection{Fraction of co-orbital types}
With the formulation of Sec.~\ref{ss:secHam}, the resonant semi-secular Hamiltonian for the motion of Earth co-orbitals is obtained by setting $p=3$ and $h=h_p=1$, thus the resonant argument is simply the difference in mean longitude between the asteroid and the Earth, i.e. $\sigma = \lambda - \lambda_3$. We are interested into the Earth co-orbital region at low eccentricity and inclination. We define the Earth co-orbital region at low eccentricity and low inclination as:
\begin{equation}
    0.994 \text{ au} < a < 1.006 \text{ au}, \quad e < 0.2, \quad I < 20 \text{ deg}.
    \label{eq:co-orbit_space}
\end{equation}
In this paper, all objects lying inside the region delimited by Eq.~\eqref{eq:co-orbit_space} are referred to as co-orbitals, even if their resonant angle circulates. These last kinds of non-resonant motions are also called \emph{sticking orbits} \citep{pan-gallardo_2025}.
There are different types of co-orbital motion in the regime of Eq.~\eqref{eq:co-orbit_space}: 1) circulation; 2) libration around $\sigma = 60$ deg or $\sigma = -60$ deg; 3) libration around $\sigma = 0$ deg with low amplitude; 4) libration around $\sigma = 180$ deg with an amplitude smaller than 180 deg; 5) libration around $\sigma = 180$ deg with an amplitude larger than 180 deg. The motion of type 2) corresponds to Trojan (also called tadpole) orbits, type 3) to quasi-satellite orbits, and type 4) to horseshoe orbits. The motion of type 5 is a mix of horseshoe and quasi-satellite (which we denote with HS+QS throughout the paper), whose existence was noticed in \citet{namouni_1999, nesvorny-etal_2002}.
The constraints of Eq.~\eqref{eq:co-orbit_space} on eccentricity and inclination are introduced to avoid node crossing with Venus, and to prevent an even more complex topology of the level curves of the Hamiltonian to appear. In fact, mixing of quasi-satellite and tadpole orbits may also appear at high enough eccentricity \citep{namouni_1999, nesvorny-etal_2002}.

Fixed the elements $(e,I,\Omega,\omega)$, orbits of a certain type are completely identified by an area in the $(a,\sigma)$-plane defined by the level curves of the semi-secular Hamiltonian $\K$. For instance, quasi-satellites are all bounded within the collision curve between quasi-satellites and horseshoes, and Trojan orbits are all bounded by the separatrix separating Trojan orbits and horseshoe orbits. 
Therefore, denote with $\mathcal{C}_k(e,i,\omega,\Omega) \subset [0 \text{ deg}, \ 360\text{ deg}] \times [0.994 \text{ au}, \ 1.006\text{ au}]$ the set corresponding to the area enclosing all the motions of type $k$, for $k=1,2,3,4,5$. We aim at computing the fraction $F_k$ of orbits of type $k$ in the Earth co-orbital region, which is obtained by:
\begin{equation}
    F_k = \frac{\displaystyle\int_e\int_{I}\int_{\omega}\int_{\Omega} \mathbf{1}_{\mathcal{C}_k} \, \text{d}\Omega \ \text{d}\omega \ \text{d}I\ \text{d}e}{\displaystyle\int_e\int_{I}\int_{\omega}\int_{\Omega} 1\,\text{d}\Omega \ \text{d}\omega \ \text{d}I \ \text{d}e},
    \label{eq:Fk}
\end{equation}
where $\mathbf{1}_{\mathcal{C}_k}$ is the indicator function of the set $\mathcal{C}_k$. Note that the fraction $F_k$ describe the instantaneous topology of the resonance and thus the geometrical availability of different type of orbits, while it does not provide information about the stability and the residence time in each state. Computing the steady-state occupation probability requires full numerical $N$-body simulations allowing state changes, as done in \citet{fenucci-etal_2026}.

\begin{figure}
    \centering
    \includegraphics[width=0.6\linewidth]{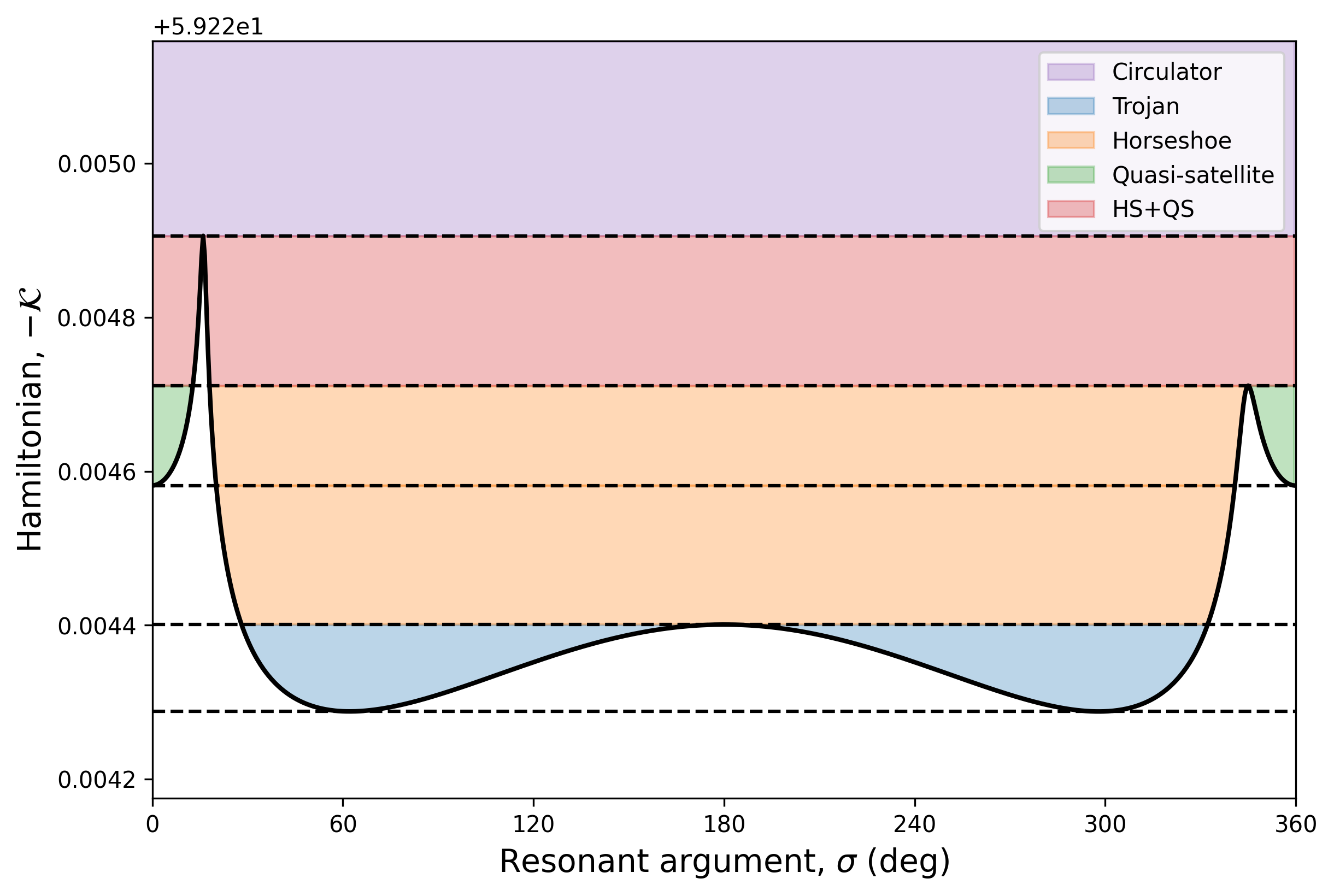}
    \caption{The Hamiltonian $-\K$ (black curve) evaluated at $a=1$ au, for $e=0.14$, $I = 10$ deg, and $\omega = 285$ deg. Different orbit types, which depend on the value of $-\K$ and the initial value of $\sigma$, are highlighted by different colors. The black dashed horizontal lines indicate the local maxima and the local minima values of $-\K$. The white area correspond to the region where the motion is forbidden.}
    \label{fig:level_curves_classification}
\end{figure}

To compute the values $F_k$ we proceed in the following way. We discretize $e,I,\omega$ with values $e_1 < \dots < e_{N_e}$, $I_1 < \dots < I_{N_I}$, $\omega_1 < \dots < \omega_{N_\omega}$ using fixed steps $\Delta e = 0.02$, $\Delta I = 1$ deg and $\Delta \omega = 5$ deg. Because planets are on circular and coplanar orbits, the Hamiltonian $\K$ does not depend on $\Omega$, and therefore this variable does not need to be discretized. 
%For each combination of $(e,I,\omega)$, we compute the value of $\K$ on a grid of $200 \times 200$ equi-spaced points in the $(a,\sigma)$ plane, from which we are able to compute the level curves of the Hamiltonian. 
For each combination of $(e,i,\omega)$, instead of precisely identifying the sets $\mathcal{C}_k$ through the computation of the separatrices and computing a 2-D integral, we proceed with a Monte Carlo method. We generate a set of $N=10~000$ random values of pairs $(a_j, \sigma_j), j=1,\dots, N$, and for each pair we compute the corresponding value of the Hamiltonian $\K_j$. The orbit type is then unequivocally determined by $\sigma_j$ and $\K_j$. In fact, the orbit type is determined by the levels of $f(\sigma) = -\K(\sigma, 1\text{ au})$ \citep{sidorenko-etal_2014}. Figure~\ref{fig:level_curves_classification} gives an example of classification of the 5 orbit types, depending on the function $f$. Let us denote with $N_k < N$ the number of level curves of type $k$ for $k=1,2,3,4,5$. Then, the value $F_k$ is approximated by
\begin{equation}
    F_k \simeq \sum_{i=1}^{N_e} \sum_{j=1}^{N_I} \sum_{l=1}^{N_{\omega}} \frac{N_k(e_i, I_j, \omega_l)}{N} .
\end{equation}
Note that each $N_k(e_i, I_j, \omega_l)$ follows a binomial distribution, thus the maximum error for the probability of each co-orbital motion is estimated to be $1/(2\sqrt{N})$. For $N=10~000$, this corresponds to a maximum of 0.5\% error.
Strictly speaking, volumes in the phase space should be computed in canonical coordinates, but as the region considered is relatively small (see Eq.~\eqref{eq:co-orbit_space}), we consider that the area ratios computed in Keplerian elements are accurate representations of the actual area ratios in the phase space.

\section{The resonant structure of the Earth co-orbital region}
\label{s:results}
\begin{figure}
    \centering
    \includegraphics[width=0.5\linewidth]{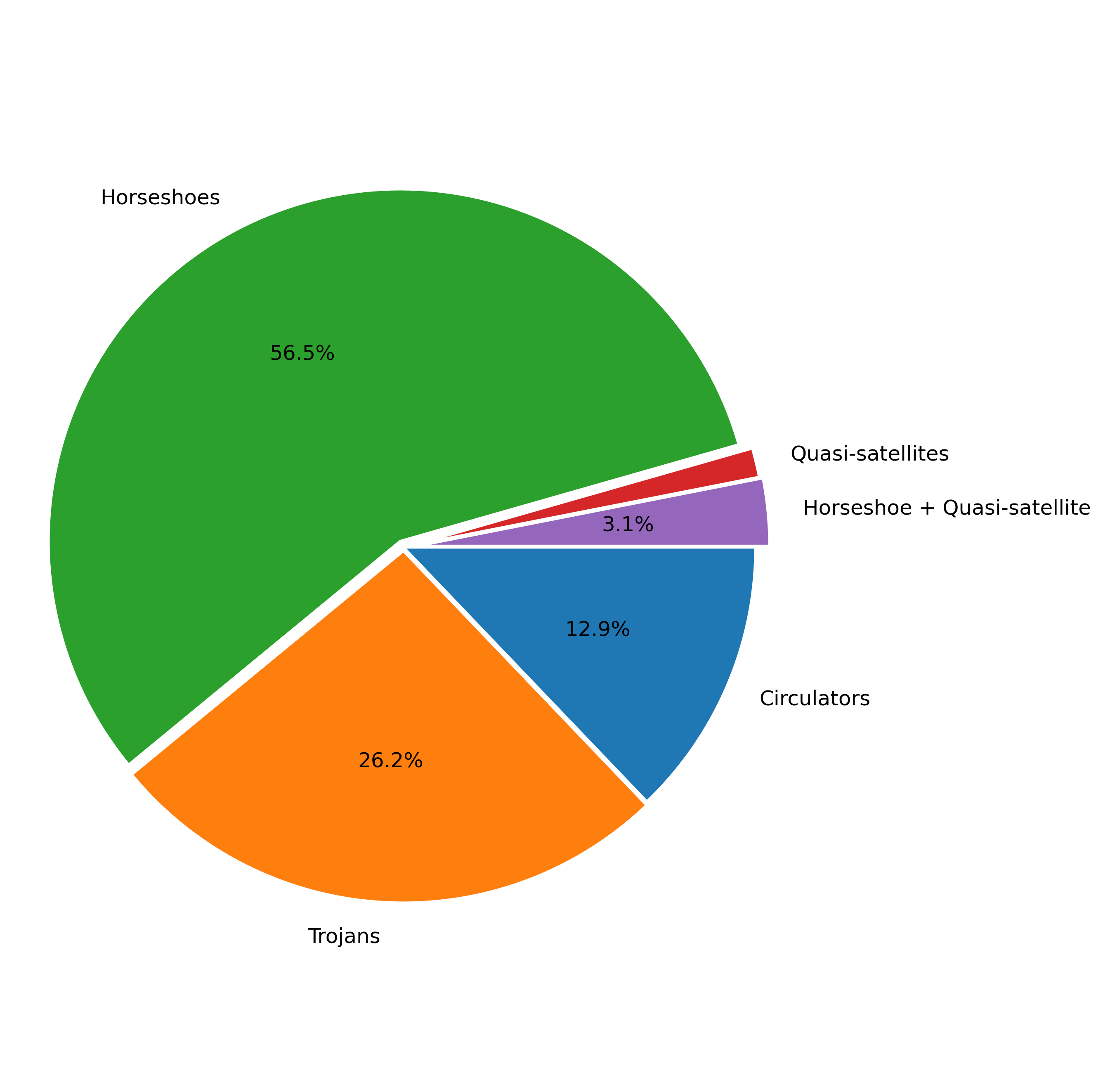}
    \caption{Fraction of horseshoe (green), quasi-satellites (red), Trojans (orange), and circulators (blue), and HS+QS (violet) within the Earth co-orbital region at low eccentricity and low inclination.}
    \label{fig:pie_chart}
\end{figure}

Figure~\ref{fig:pie_chart} shows the fraction of each orbit type as obtained with the method described above. The Earth co-orbital region is vastly dominated by horseshoe orbits, which fill more than half of the phase space. The second largest type is represented by Trojan orbits, which comprise about 26\% of the phase space. The mixed class HS+QS appears in the space defined by Eq.~\eqref{eq:co-orbit_space}, and it represents instead a total of about 3\% of the phase space. Circulators represent about 13\% of the phase space, while quasi-satellite orbits are the rarest one, filling only about 1.33\% of the volume. Other mixes classes, such as quasi-satellite and Trojan, were not found on the other hand.
The fact that the Earth co-orbital region is mostly filled by horseshoe orbits has an analytical explanation. Previous studies about the restricted three-body problem showed that the width of the region occupied by Trojan orbits is proportional to $(m_2/m_1)^{1/2}$ \citep{dermott-murray_1981}, while that occupied by horseshoe orbits to $(m_2/m_1)^{1/3}$ \citep{robutel-pousse_2013}, where $m_1$ is the mass of the primary and $m_2$ that of the secondary. Thus, the co-orbital region of terrestrial planets is mostly filled by horseshoe orbits.
Note also that the percentage of circulators is determined by the limits in semi-major axis of Eq.~\eqref{eq:co-orbit_space}, however we are still interested in their number to understand how the phase space volume occupied by the resonance varies as a function of the other orbital elements. 

\begin{figure}
    \centering
    \includegraphics[width=0.95\textwidth]{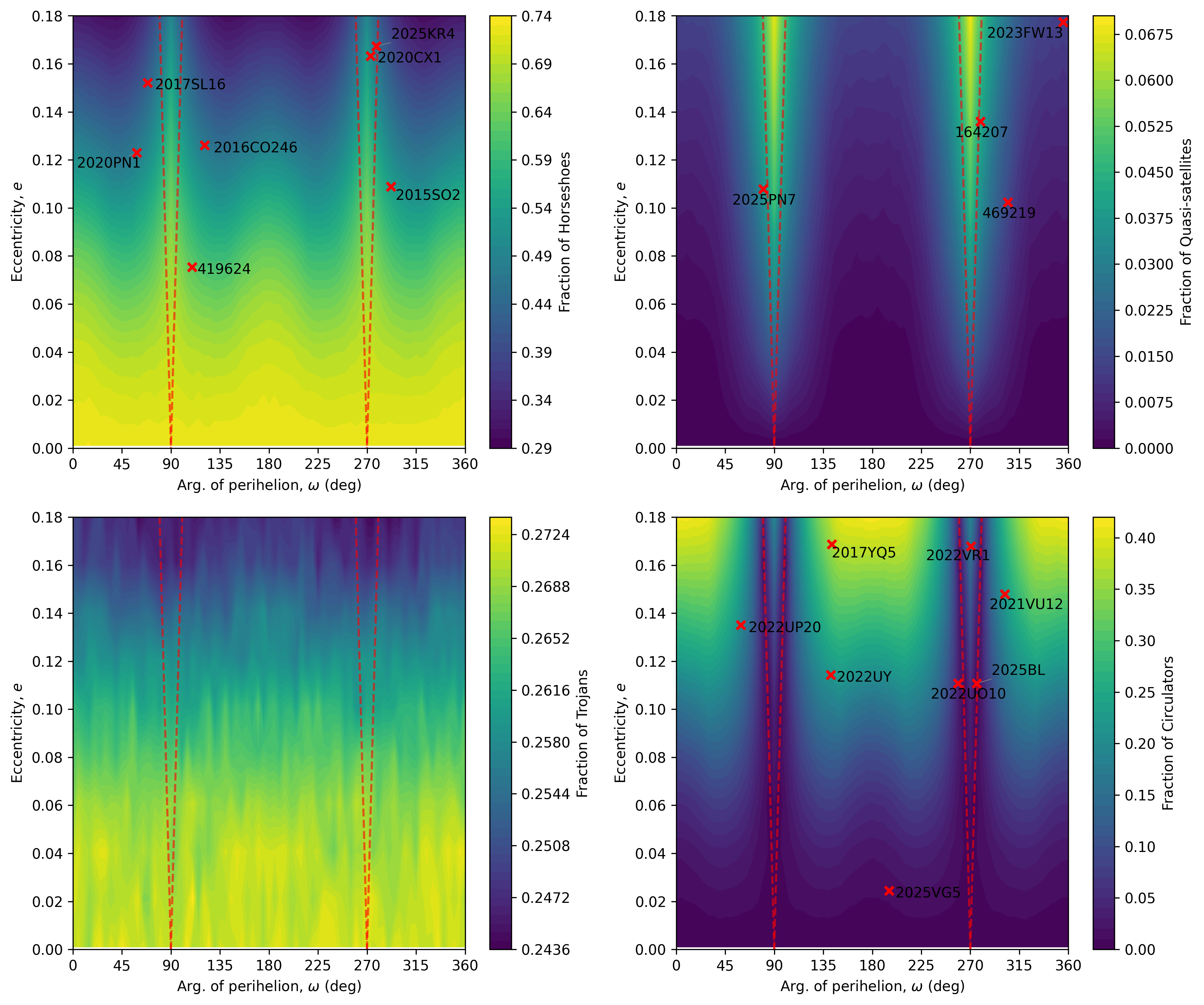}
    \caption{Fraction of horseshoe (top left panel), of quasi-satellites (top right panel), of Trojans (bottom left panel), and of circulators (bottom right panel)  in the $(\omega, e)$ plane, obtained by integrating only over $\Omega$ and $I$. The red dashed lines represent the node crossing curves of Eq.~\eqref{eq:cen_coll_cond} computed for $a=1$ au.}
    \label{fig:frac_w_e}
\end{figure}

\begin{figure}
    \centering
    \includegraphics[width=0.95\textwidth]{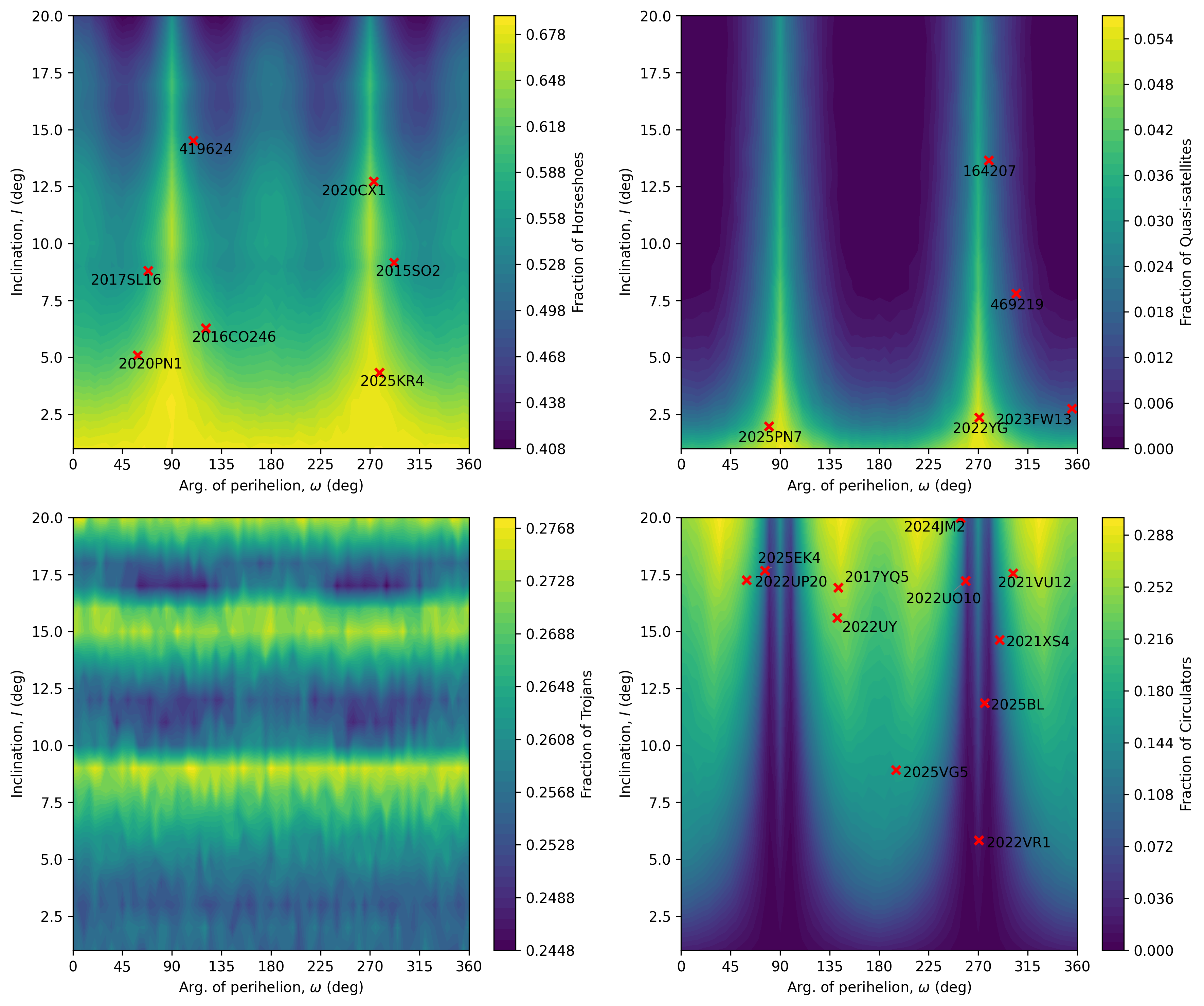}
    \caption{Fraction of horseshoe (top left panel), of quasi-satellites (top right panel), of Trojans (bottom left panel), and of circulators (bottom right panel) in the $(\omega, I)$ plane, obtained by integrating only over $\Omega$ and $e$.}
    \label{fig:frac_w_i}
\end{figure}

\begin{figure}
    \centering
    \includegraphics[width=0.95\textwidth]{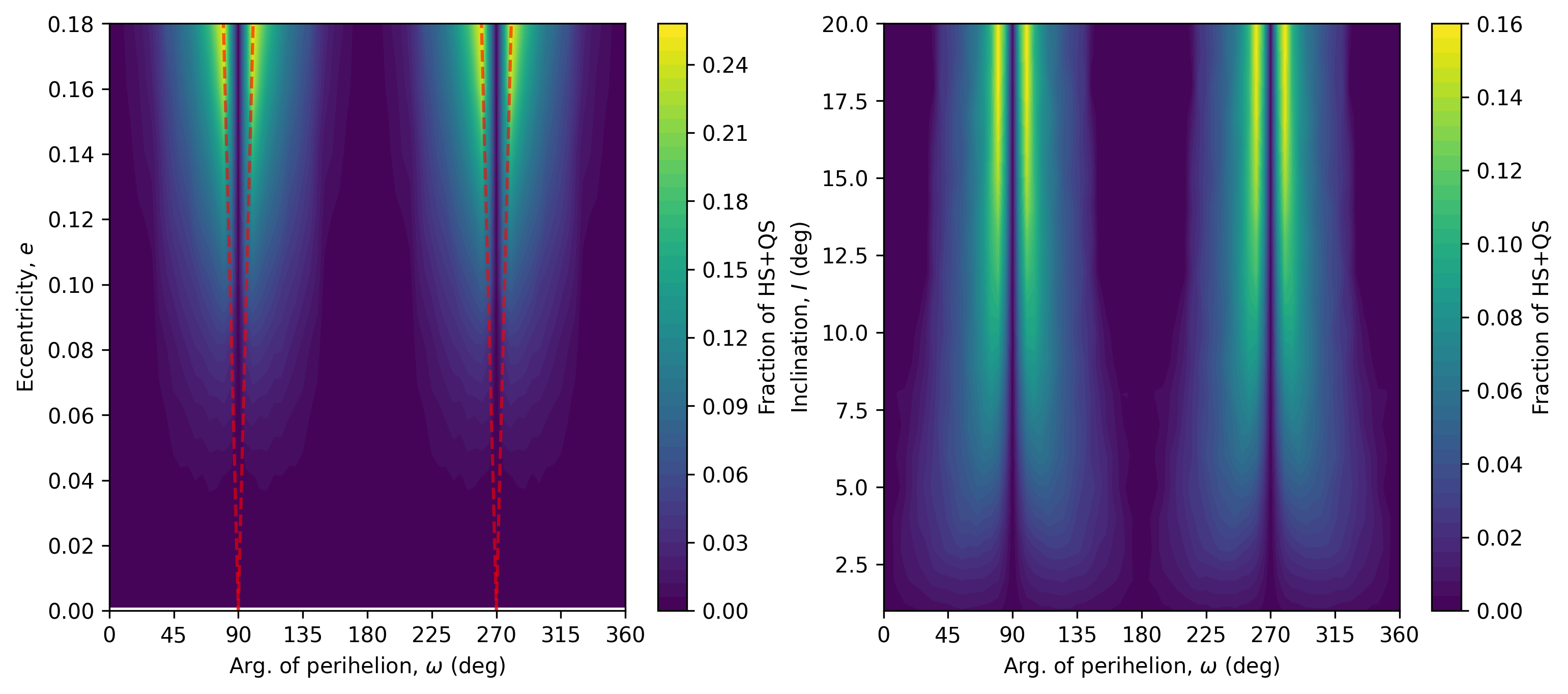}
    \caption{Fraction of mixed HS+QS type in the $(\omega, e)$ plane (left panel), and in the the $(\omega, I)$ plane (right panel).}
    \label{fig:frac_hsqs}
\end{figure}

To identify possible inhomogeneities in the phase, the integrals in Eq.~\eqref{eq:Fk} can be computed only over $\Omega$ and another one of the orbital elements among $e,I,\omega$. This permits to analyze the fraction of each orbit type in the plane of the two remaining orbital elements. Figure~\ref{fig:frac_w_e} and Fig.~\ref{fig:frac_hsqs}, left panel, show the results obtained by integrating only over the inclination $I$, thus the fractions are presented in the $(\omega, e)$ plane. Figure~\ref{fig:frac_w_i} and Fig.~\ref{fig:frac_hsqs}, right panel, show the results obtained by integrating only over the eccentricity $e$, and the fractions are represented in the $(\omega, I)$ plane. It is worth noting that the occurrence of horseshoe, quasi-satellite, HS+QS, and circulator orbits shows evident inhomogeneities through the phase space, while the fraction of Trojan orbits shows small variations, being always between roughly 24\% and 27\%, with the lowest concentration appearing at higher eccentricity.  

% Horseshoe orbits
Horseshoe orbits occupy more than 50\% of the phase space at $e<0.1$. This percentage decreases for larger eccentricity values, and it shows a general dependency on the argument of perihelion $\omega$. In fact, at $e\geq0.1$ horseshoe orbits are more frequent in the neighborhood of $\omega = 90, \ 270$ deg, and they reach a percentage as low as 30\% near $\omega = 45,135,225,315$ deg and $e>0.15$. A similar dependency on $\omega$ is present also in the percentage of horseshoes in the $(\omega, I)$ plane. In addition, horseshoe orbits are more frequent at low inclination values, with two high frequency areas even at high inclination near $\omega = 90, \ 270$ deg.
Note that the node crossing with Earth happens when the eccentricity and the argument of perihelion satisfy the equation
\begin{equation}
    \cos \omega = \pm{\frac{a(1-e^2) - 1}{e}}.
    \label{eq:cen_coll_cond}
\end{equation}
The node crossing curves for $a=1$ au are reported in Fig.~\ref{fig:frac_w_e} as red dashed lines, and they are placed around $\omega = 90, 270$ deg. The secular term $\mathcal{K}_{\text{sec}}$ is divergent at the crossing configuration for a specific value of $\sigma$ \citep{fenucci-etal_2022}, thus giving rise to an unbound maximum in the Hamiltonian $\K$ \citep{sidorenko-etal_2014}. This behavior of the Hamiltonian thus allows horseshoes orbits to always appear at node crossing, and be more frequent also close to this configuration. 

% Quasi-satellite orbits
Quasi-satellite orbits are, on the other hand, relatively rare in the whole phase space investigated here, never exceeding $\sim$6\% of the total volume. A strong dependency on the argument of perihelion is also present, similarly to the fraction of horseshoe orbits. At eccentricity $e<0.04$ quasi-satellites are very rare, and also near $\omega = 180$ deg at any eccentricity value. The peak in the volume occupation of quasi-satellites orbits is reached near the crossing singularity, showing two clear higher density areas. Quasi-satellites orbits are mostly present near the crossing singularity because of the appearance of a local minimum of $f$ at $\sigma=0$ deg. In fact, $\sigma = 180$ deg is always a maxima, and two more appear near the singularity. Since there are always two minima near $\sigma = \pm 60$ deg, and the number of maxima and minima should be equal because of the Euler characteristics of the circle, another local minima exists, thus allowing a small island where the quasi-satellite motion is possible. A detailed characterization of appearance of quasi-satellite orbits is presented also in \citet{sidorenko-etal_2014}.

The projection in the $(\omega, I)$ plane shows that quasi-satellites orbits are most frequent at inclination $I<5$ deg, with two concentration areas persisting at higher inclination around $\omega=90, \ 270$ deg. Note that asteroids experiencing quasi-satellite and horseshoe states are also often coupled with a libration of the argument of perihelion $\omega$ around these two values, as for the cases of (469219) Kamo`oalewa and 2015~SO$_2$ \citep{delafuente-delafuente_2016, delafuente-delafuente_2016b}.
The libration of $\omega$ has been explained with secular models for the evolution of $e,I,\omega, \Omega$ using an adiabatic approximation \citep{namouni_1999, nesvorny-etal_2002, fenucci-etal_2022}, through which Kozai-type secular evolution but in a resonance framework \citep{sidorenko-etal_2014, saillenfest-etal_2016} is found. 

The mixed class HS+QS also appears only near the node crossing singularity (see Fig.~\ref{fig:frac_hsqs}), with percentages of occupancy up to $\sim$20\% at eccentricity larger than 0.12, or inclination larger than 10 deg. The occurrence of this mix class is again due to the singularity. In fact, this type of motion is possible only when two different finite maxima (other than that at $\sigma = 180$ deg), or a maximum and a singularity, appear in the function $f$, as it is shown in the example of Fig.~\ref{fig:level_curves_classification}.

% Trojan orbits
For Trojan orbits, the fact that their fraction remains nearly constant throughout the explored phase space is driven by the topological structure of the 1:1 resonance. Trojan islands exist for all combinations of $(e,I,\omega)$ values because they are associated to the absolute minima of $f$, which always exist, and the area enclosing all these orbits has small variations, at least at small eccentricity and low inclination. In addition, the homogeneity in $\omega$ is explained by the fact that conjunction with Earth is not possible for these orbits, thus the perihelion orientation is almost irrelevant for Trojan dynamics. 

% Circulators
Circulator orbits on the other hand show a pattern complementary to horseshoe, quasi-satellite, and HS+QS orbits. Circulators fill a smaller portion of the co-orbital region in Eq.~\eqref{eq:co-orbit_space} at eccentricity values lower than 0.1 and at inclination smaller than 2.5 deg, reaching a frequency as low as 3\%. On the contrary, the largest volume occupation appears at high eccentricity and high inclination, especially near $\omega = 0, \ 180$ deg, where they fill a percentage of the phase space between 20\% and 30\%. Low frequency gaps are also present around node crossing with the Earth, which is where where horseshoe, quasi-satellites, and HS+QS orbits are more frequent. The increasing fraction of circulators for growing eccentricity is a consequence of the fact that the resonance width decreases as the eccentricity increases, thus increasing the volume filled by this type of orbits.  

\section{Short-term stability maps of the Earth co-orbital region}
\label{s:megno}
The distributions of Fig.~\ref{fig:frac_w_e} may hint about possible stable and unstable regions of the Earth co-orbital space. In fact, the intersection between the orbit of the asteroid and that of the Earth generally happens at node crossings, defined by Eq.~\eqref{eq:cen_coll_cond}, thus indicating which co-orbital type may be more chaotic than others.
However, full numerical $N$-body propagations are needed to properly study stability properties of Earth co-orbitals. To this purpose, we performed full $N$-body integrations of Earth co-orbitals using the IAS15 integrator \citep{rein-spiegel_2015} included in the \texttt{REBOUND}\footnote{\url{https://rebound.hanno-rein.de/}} package \citep{rein_2012}. 
In the simulations, we took into account the gravitational perturbations of the Sun and all the planets from Venus to Neptune. Initial conditions for planets were taken from the JPL Horizons\footnote{\url{https://ssd.jpl.nasa.gov/horizons/}} system. Random orbital elements for Earth co-orbital asteroids were generated with uniform distributions, and using the constraints of Eq.~\eqref{eq:co-orbit_space}.

To evaluate the stability of an asteroid, we use the Mean Exponential Growth factor of Nearby Orbits (MEGNO) chaos indicator \citep{cincotta-simo_2000, cincotta-simo_2003}, computed after 1000 yr of orbital evolution. The computation of the MEGNO indicator is implemented in the \texttt{REBOUND} package. This timespan is chosen so that the slow-evolving orbital elements $e,I,\omega$ and $\Omega$ do not suffer much from secular perturbations, thus these objects stay in a regime where the averaged model presented in Sec.~\ref{s:methods} is generally valid. At the same time, the timespan is long enough for objects to enable state switches between different co-orbital states, and possibly undergo close encounters with Earth. 
In fact, the characteristic time between state switches was indicated by \citet{sidorenko-etal_2014}. At low eccentricity and inclination, this is approximated as $\tau_{\text{QS}} = (e^2 + i^2)^{3/2}/(2\pi \mu)$ for quasi-satellites, and as $\tau = (e^2+i^2)/(2\pi\mu)$ for other co-orbital types, where $\mu = 3.0035 \times 10^{-6}$ is the mass parameter of the Earth and the time is expressed in years. Typical residence times in a quasi-satellite state is therefore of the order of $\sim$200--1000 years, while they can be slightly longer for other co-orbital types.

% Description of MEGNO
We recall that the MEGNO indicator converges to the value of 2 for a quasi-periodic orbit. On the other hand, in the case of chaotic orbits, the MEGNO indicator grows linearly with time, with a slope proportional to the maximum Lyapunov exponent. Recall also that, for a MEGNO larger than 2, the Lyapunov time of the orbit can be estimated as $T/(2 \times \textrm{MEGNO})$, where $T$ denotes the total integration timespan. Therefore, the MEGNO indicator is an efficient tool to distinguish between regular and chaotic motions. 

\begin{figure}[!ht]
    \centering
    \includegraphics[width=0.98\textwidth]{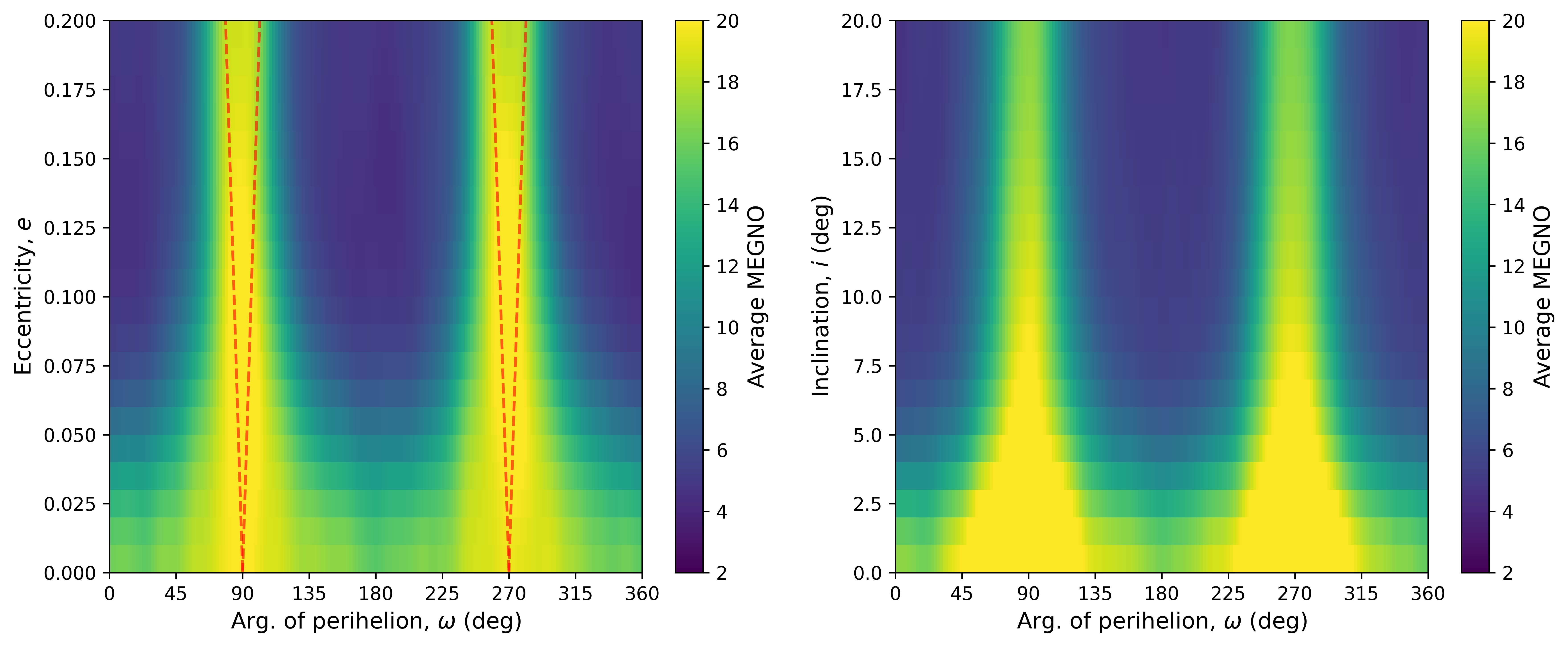}
    \includegraphics[width=0.98\textwidth]{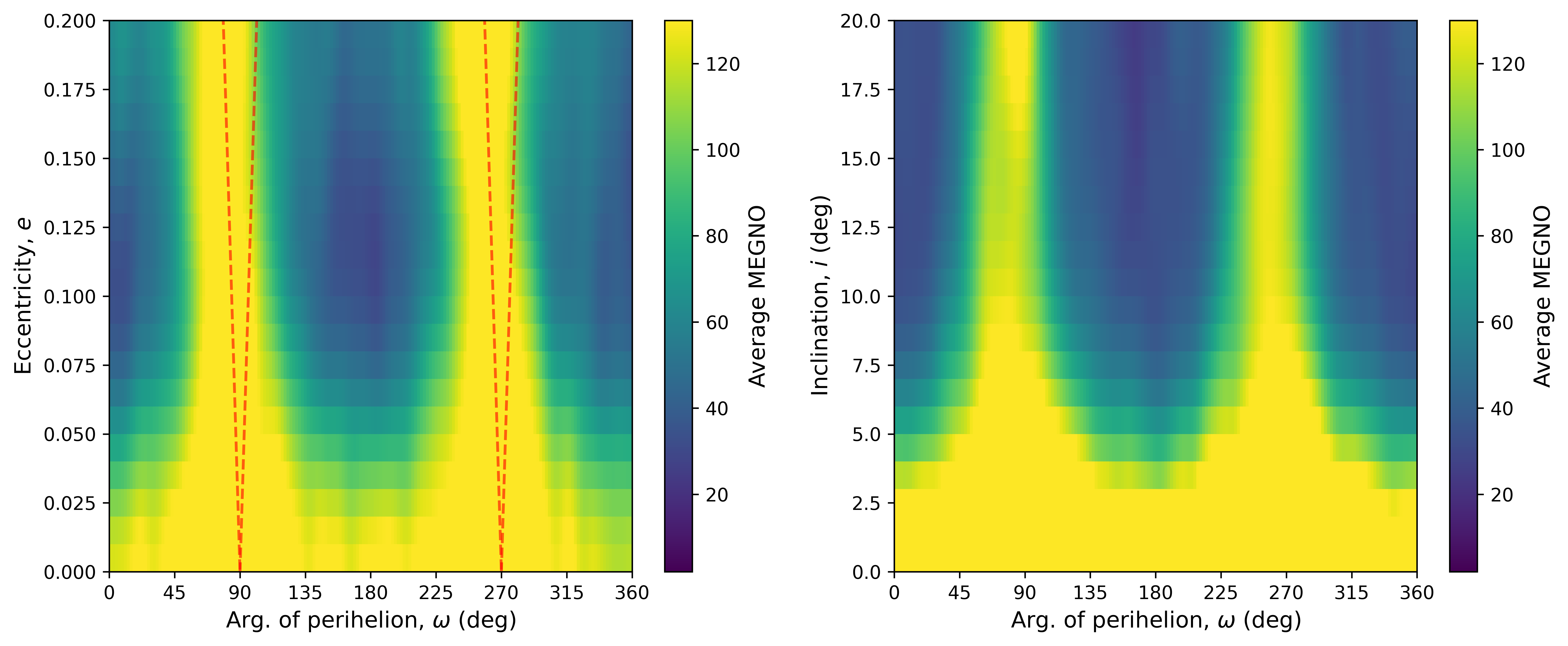}
    \caption{Maps of the average MEGNO indicator in the plane $(\omega, e)$ (left panel), and in the plane $(\omega, I)$, obtained by integrating 200~000 randomly generated Earth co-orbitals. The red dashed lines represent the node crossing curves of Eq.~\eqref{eq:cen_coll_cond} computed for $a=1$ au. The upper figures are obtained with an integration of 1000 yr, while the lower panel with an integration over 10~000 yr.}
    \label{fig:megno_map_1000yr}
\end{figure}

We computed the MEGNO indicator for a sample of 200~000 randomly generated Earth co-orbitals. We then computed the average MEGNO indicator in bins in the planes of $(\omega, e)$ and $(\omega, I)$, so that it is possible to compare them to the results of Fig.~\ref{fig:frac_w_e}. The top panels of Fig.~\ref{fig:megno_map_1000yr} show the maps of the average MEGNO for an integration of 1000 yr. Regions with high average MEGNO values are present in both projections in $(\omega, e)$ and in $(\omega, I)$, clustered in particular at low inclination and low eccentricity. In addition, two other regions with high average MEGNO are aligned along $\omega = 90, 270$ deg, similar to those seen in Fig.~\ref{fig:frac_w_e} and \ref{fig:frac_w_i}. The high MEGNO values seen in these areas are caused by close encounters with the Earth. In fact, the node crossing with the orbit of the Earth is aligned along $\omega=90, 270$ deg, while orbits with low inclination or low eccentricity are more statistically subject to suffer from close encounters with Earth, since they have a generally small distance from the orbit of the Earth along the whole orbit. Note that in these last cases, the minimum orbit intersection distance (MOID) may be away from the node crossing. Other long-term sources of chaotic behaviors, such as secular perturbations or secular resonances \citep{qi-qiao_2022}, do not have enough time to act over a timespan of 1000 yr. For longer integration timespans, the secular evolution of the perihelion due to the secular dynamics could slightly change the qualitative stability picture, because some orbits may be pushed towards the Earth node crossing over a long timescale. In particular, the high chaotic region would become slightly larger over values slightly smaller than 90 deg and 270 deg, because positive $\omega$ circulators are more common than negative ones \citep{namouni_1999}. This effect can be already in an integration that we extended for a total time of 10~000 yr, shown in the bottom panels of Fig.~\ref{fig:megno_map_1000yr}.

On the other hand, regions at eccentricity lager than 0.05 and inclination larger than 7.5 deg, and around $\omega = 0, 180$ deg, have a generally low average MEGNO value, indicating that objects lying in these portions of the phase space have longer stability times. This is explained by the fact that they are far from the node crossing with Earth, which can be eventually reached only over longer timespans due to the precession of $\omega$. The results presented above were obtained with a dynamical model which did not include the Moon among the perturbing bodies. Including the Moon in the model would add another body for which close encounters are possible, therefore increasing the chances for chaotic behavior. However, close approaches with the Moon would necessarily still happen either near the node crossing with the Earth or, more in general, near a low Earth MOID. Thus, the stability maps presented in Fig.~\ref{fig:megno_map_1000yr} would not be affected qualitatively.

\begin{figure}
    \centering
    \includegraphics[width=0.98\textwidth]{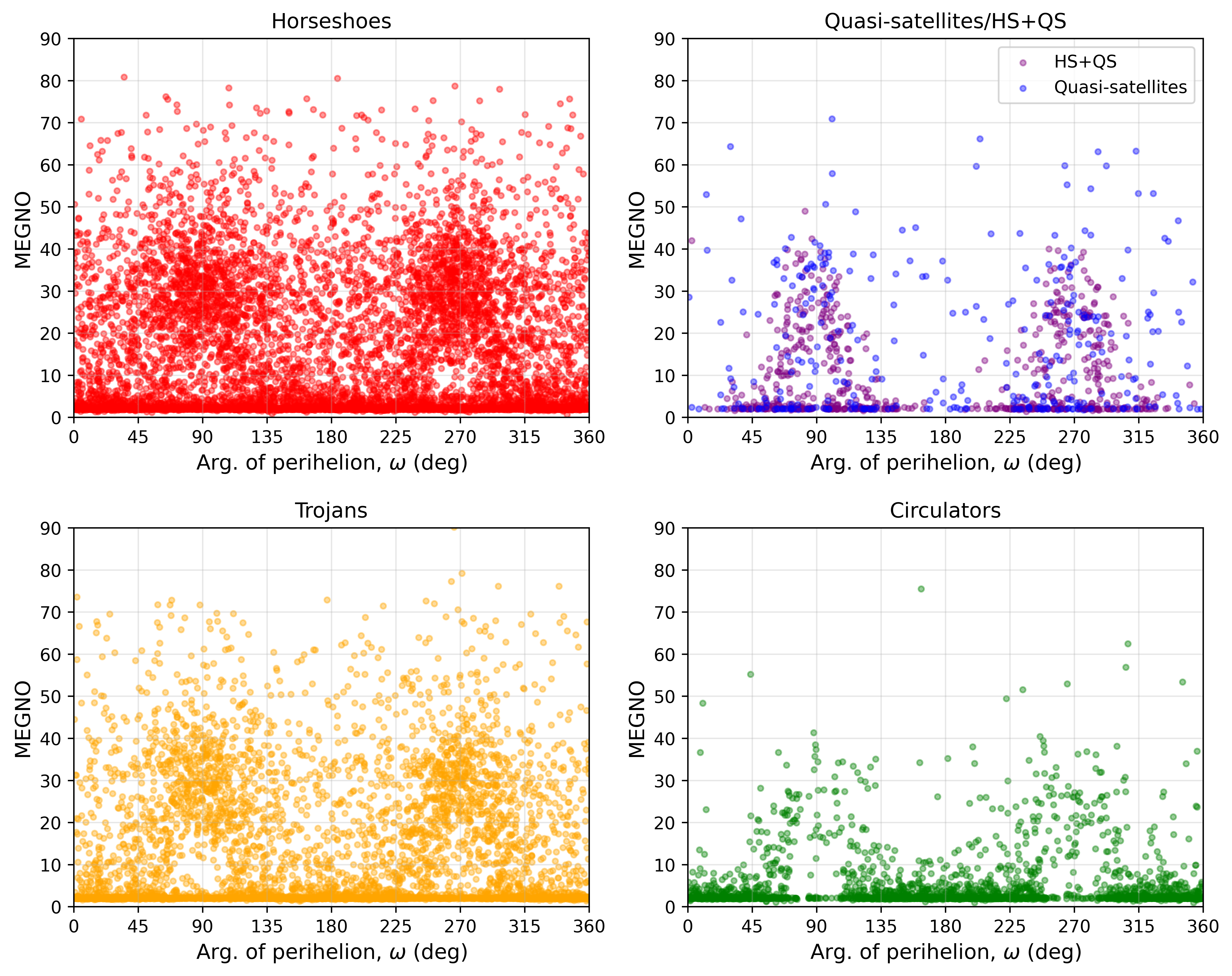}
    \caption{Distribution of the MEGNO value as a function of $\omega$, for different co-orbital types. The co-orbital type is reported above each panel, and it refers to the initial state identified by the corresponding initial level of $f$.}
    \label{fig:w_megno_classification}
\end{figure}

Note that horseshoes and Trojans contribute the most to the average stability map of Fig.~\ref{fig:megno_map_1000yr}, since they are the most represented orbital type, computed with the 1000 yr integration. To understand the stability properties of different co-orbital states, we extracted $20~000$ objects from the above simulations, and computed the initial classification using the method described in Sec.\ref{s:methods}. Figure~\ref{fig:w_megno_classification} shows the scatter plot of MEGNO vs. argument of perihelion $\omega$, for different orbital classes. Horseshoe and Trojan orbits appear to be the most chaotic ones, followed by circulators. Some quasi-satellites also appear to be unstable over 1000 yr, but with a smaller fraction than the other three classes. Recall that a MEGNO of 2 corresponds to an orbit stable over the a long timespan, and many of them can be seen in the plots. On the contrary, the stability of orbits with a larger MEGNO can be interpreted in terms of their Lyapunov time. For a 1000 yr integration, a MEGNO of 5, 10, 20, and 30 correspond to a Lyapunov time of 100 yr, 50 yr, 33 yr, 25 yr, and 17 years, respectively. The plots of Fig.~\ref{fig:w_megno_classification} show that objects placed in a Earth-crossing orbit may have an extremely short Lyapunov time. Note also that, due to gravitational perturbations, the initial state may not be kept for the whole integration timespan. However, these results still give information about which co-orbital type of motion tend to be more chaotic in the long term.  

The stability of Earth co-orbitals over a longer timescale would need to be evaluated in other terms than the MEGNO indicator. \citet{dvorak-etal_2012} evaluated the stability of Earth Trojans over timescales of $\sim10^7$ yr in terms of the maximum eccentricity reached by test particles, and provided maps in the plane of semi-major axis and inclination. Their results were later extended by \citet{zhou-etal_2019}, who used spectral analysis to evaluate the stability over 12 Myr. The authors also mapped the locations of secular resonances inside the Earth co-orbital region, showing that they also contribute to the instability regions seen in their maps. Moreover, the influence of the Yarkovsky effect was also investigated, finding that the survival time decreases when this effect is included in the model. This was also found in a dedicated study about the long-term dynamics of the quasi-satellite Kamo`oalewa \citep{fenucci-novakovic_2021}.

\section{Discussion}
\label{s:discussion}
\subsection{Earth impact risk}
% Dynamical protection from close encoutners
The results presented in Fig.~\ref{fig:frac_w_e}, \ref{fig:frac_w_i}, \ref{fig:frac_hsqs}, and \ref{fig:megno_map_1000yr} provide a view of how short-term impactors are related to different Earth co-orbital types. 
For a collision to happen, two necessary and sufficient conditions need to be satisfied: 
1) the MOID with Earth needs to be 0, and 
2) the Earth and the asteroid should be at the same point along their orbits. Condition 2) is satisfied only for one specific value $\sigma=\sigma_c$ of the resonant argument. For circular orbits, the collision angle corresponds to exact conjunction $\sigma_c = 0$ deg, while for low eccentricity orbits we have $\sigma_c = \mathcal{O}(2e)$ (see Appendix~\ref{app:collision_angle}), meaning that collisions can still happen only near conjunction. 

The node crossing condition of Eq.~\eqref{eq:cen_coll_cond} also implies that the MOID is 0, thus it already indicates a portion of phase space where collisions can happen. 
The relative longitude $\sigma$ has a different behavior for each different type of co-orbital motion. In the averaged system, $\sigma$ librates around 180 deg with an amplitude of almost 180 deg for horseshoe orbits, thus they stay most of the time away from Earth and never reach a precise conjunction. Quasi-satellite orbits are characterized by a libration of $\sigma$ around 0 deg with low amplitude, but always staying safely away from Earth. Trojan orbits at small eccentricity and low inclination never get close to $\sigma=0$ deg, thus they are naturally out of collision risk. On the other hand, circulators can cross the conjunction condition $\sigma=0$ deg, as well as the mixed HS+QS class.

When propagated in a full $N$-body setting, however, all the co-orbital types show chaotic behavior over a relatively short timescale of 1000 yr, caused by close approaches with Earth. This is seen in the high MEGNO values displayed in the distributions of Fig.~\ref{fig:megno_map_1000yr}. There are two main mechanisms which cause Earth co-orbital to undergo close approaches in such short timescale. First, in the averaged model, small oscillations of $\sigma$ due to the gravitational perturbation of other planets are averaged out. In a full $N$-body dynamics however, these perturbations can cause a horseshoe orbit to shortly cross the $\sigma =0$ deg value, thus being possibly subject to close encounters. This type of behavior is seen also some known Earth co-orbital (see Sec.~\ref{ss:knownpop}).  
Second, a specific co-orbital state is often temporary. If the initial orbit is close to a separatrix, even small changes in $e,I,\omega, \Omega$ can modify the instantaneous level curves of $\K$ so that the adiabatic invariant can not be met by the same initial co-orbital state, thus forcing the object to switch to another state. This process can happen multiple times even in a short timescale of 1000 yr, as seen for example in the motion of asteroids (469219) Kamo`oalewa, 2022~VR1, or 2015~SO2 \citep[see][]{fenucci-etal_2026}. Thus, co-orbital state switches due to the adiabatic dynamics can cause late impactors for orbits that initially did not satisfy the conjunction condition. These may be seen on a timescale longer than 100 years, which is the timespan usually used in impact monitoring systems such as Sentry\footnote{\url{https://cneos.jpl.nasa.gov/sentry/}} by NASA or Aegis by ESA \citep{fenucci-etal_2024b}. From a planetary defence point of view it is therefore important to discover the whole population of Earth co-orbitals, and not only the subset that satisfies the MOID and the conjunction conditions.

\begin{table}[!ht]
    \centering
    \begin{tabular}{cccrrcccc}
    \hline
\hline
Desig. & $a$ (au) & $e$ & $I$ (deg) & $\omega$ (deg) & $H$ (mag) & Orbit type & $\min_d$ (au) & Year\\
\hline
(164207) Cardea&1.001122&0.135938  &  13.6529    & 279.3410 &21.11& QS &0.147&1967\\
2015~SO2&0.994150&0.108850&9.1675                & 291.8806 &23.70& HS &0.036&2015\\
2016~CO246&0.994773&0.126172&6.2947              & 120.9397 &25.62& HS &0.039&2018\\
2017~SL16&0.995837&0.152046&8.8030               & 68.4632 &25.97 & HS &0.020&2019\\
2017~YQ5&0.995763&0.168715&16.9247               & 142.7842 &24.72& C  &0.121&2018\\
2020~CX1&1.004216&0.163249&12.7321               & 273.2899 &24.08& HS &0.012&2021\\
2020~PN1&1.005294&0.123008&5.0901                & 58.7961 &25.39 & HS &0.024&2021\\
2021~VU12&1.004927&0.147979&17.5482              & 301.5958 &24.18& C  &0.070&2022\\
2021~XS4&0.995861&0.189525&14.6254               & 289.2652 &25.86& C  &0.031&2021\\
2022~UO10&1.005478&0.110736&17.2227              & 258.6020 &26.69& C  &0.019&2023\\
2022~UP20&1.002200&0.135075&17.2497              & 59.3061 &25.21 & C  &0.057&2025\\
2022~UY&0.996913&0.114363&15.5967                & 141.7725 &25.45& C  &0.080&2023\\
2022~YG&1.003685&0.196207&2.3669                 & 270.9303 &26.69& QS &0.013&2061\\
2022~VR1&  0.998234&0.168152         &     5.7826& 270.5657 &24.74& C  &0.010&2022\\
2023~FW13&0.999365&0.177395&2.7512               & 355.0946 &25.95& QS &0.063&1993\\
2024~JM2&0.994505&0.199786&19.9786               & 253.9051 &23.82& C  &0.017&2024\\
2025~BL     &1.000405&0.110752&11.8515           & 275.8257 &25.29& C  &0.008&2014\\
2025~EK4    &0.994498&0.193695&17.6788           & 75.9682 &24.01 & C  &0.013&2118\\
2025~KR4    &1.003767&0.167396&4.3341            & 278.5448 &25.95& HS &0.007&2114\\
2025~PN7    &1.000990&0.107930&1.9755            & 79.8882 &26.39 & QS &0.027&1980\\
2025~VG5    &1.003516&0.024513&8.9180            & 195.3210 &26.79& C  &0.028&2023\\
(419624) 2010~SO16   &1.003943&0.075415& 14.5134 & 109.2590 &20.46& HS &0.130&2010\\
(469219) Kamo`oalewa&1.000976&0.102368 & 7.8031  & 304.5481 &24.00& QS &0.098&1987\\
\hline
    \end{tabular}
    \caption{List of known Earth co-orbitals extracted from the ESA NEOCC database on 16 January 2026. The first column is the designation of the asteroid, followed by the semi-major axis $a$, the eccentricity $e$, the inclination $I$, the argument of perihelion $\omega$, and the absolute magnitude $H$. The column Orbit type indicates the type of motion, where HS = horseshoe, QS = quasi-satellite, and C = circulator. The columns $\min_d$ and Year indicate the closest close approach with Earth from 1 January 1950 to 17 January 2126 and the corresponding year, respectively. }
    \label{tab:known-coorbitals}
\end{table}

\subsection{The known population of Earth co-orbitals}
\label{ss:knownpop}
% Comparison with known Earth co-orbitals
To check how the results of Sec.~\ref{s:results} compare to the known population of Earth co-orbitals, we retrieved data from the NEO Coordination Centre\footnote{\url{https://neo.ssa.esa.int/}} (NEOCC) database by the European Space Agency (ESA), on 16 January 2026. We selected all the asteroids fulfilling the conditions of Eq.~\eqref{eq:co-orbit_space}, using a maximum absolute magnitude value of 27. A total of 22 NEAs were found, and are reported in Table~\ref{tab:known-coorbitals} together with their corresponding nominal orbital elements, absolute magnitude and orbit type. Their orbit type is reported in the table, and it was determined by analyzing the time evolution of the resonant angle $\sigma$. 
Note that there are currently only two known Earth Trojans: 2010~TK7 \citep{connors-etal_2011} and 2020~XL5 \citep{hui-etal_2021, santana-ros-etal_2022}. However, they are not included in Tab.~\ref{tab:known-coorbitals} because they do not fulfill the constraints of Eq.~\eqref{eq:co-orbit_space}, since the value of the inclination of 2010~TK7 is larger than 20 deg and the eccentricity of 2020~XL5 is larger than 0.2. 

The resonant angle of the objects of Tab.~\ref{tab:known-coorbitals} was obtained by integrating the nominal orbit of the asteroids and the orbit of the Earth with the high-fidelity dynamical model implemented in the ESA Aegis software \citep{fenucci-etal_2024b}. The closest close approach with Earth, truncated to the third decimal digit, from 1 January 1950 to 17 January 2126 is also reported in the column $\min_d$, with the corresponding year of the event. Note that none of the asteroids in the table is in the Risk List, meaning that they have zero probability of impacting the Earth in the next 100 years.
The location of these NEAs in the $(\omega, e)$ and in the $(\omega,I)$ planes is marked by red crosses in Fig.~\ref{fig:frac_w_e} and \ref{fig:frac_w_i}, respectively. 

The known Earth co-orbitals fit within the regions where this kind of motion occupies the most space, with few exceptions. Asteroid 2022~VR1 is currently a circulator, and it is located in a volume of the phase space where circulators are quite rare. However, full numerical simulations of the future evolution show that it will transition to a horseshoe orbit in about 80 years, thus well fitting in a volume where this type of motion is more likely. 
Note that, among the known co-orbitals, only two of them reach a close approach with the Earth with a distance smaller than 0.01 au: 2025~BL and 2025~EK4. The first asteroid is a circulator, lying exactly on the node crossing lines showed in Fig.~\ref{fig:frac_w_e}, which reached conjunction in 2014. The second one is placed on a horseshoe orbit, which will reach conjunction in 2114 because of small gravitational perturbations.

\subsection{The expected population of Earth co-orbitals}
% Comparison with NEOMOD3
The fractions computed in Sec.~\ref{s:results} and showed in Fig.~\ref{fig:pie_chart} can be combined with NEA population models to get an estimate of the number of NEAs in each co-orbital type. To this purpose, we used the NEOMOD3 model by \citet{nesvorny-etal_2024b}, which gives an orbital and absolute magnitude distribution of the NEA population. By using the tool provided with the model, we generated a population of NEAs with diameter between 10 m and 100 m, which comprises a total of about 6 million objects. From this populations, we extracted only those in the Earth co-orbital region, using the constraints of Eq.~\eqref{eq:co-orbit_space}. 
Recall that the argument of perihelion $\omega$ is uniformly distributed for NEA population models. For each orbit of the model, we determined which point of the grid in $e,I,\omega$ is the closest, and recovered the fraction of each orbital type in that specific snapshot of the phase space. By summing over all the objects of the model, we get the estimated number of each co-orbital type, which takes into account also the non-uniform distribution of eccentricity and inclination of the NEA population models. In addition, we assume a Poisson statistics on the distribution, therefore the standard deviation is simply the square root of the corresponding number.

\begin{figure}
    \centering
    \includegraphics[width=0.9\textwidth]{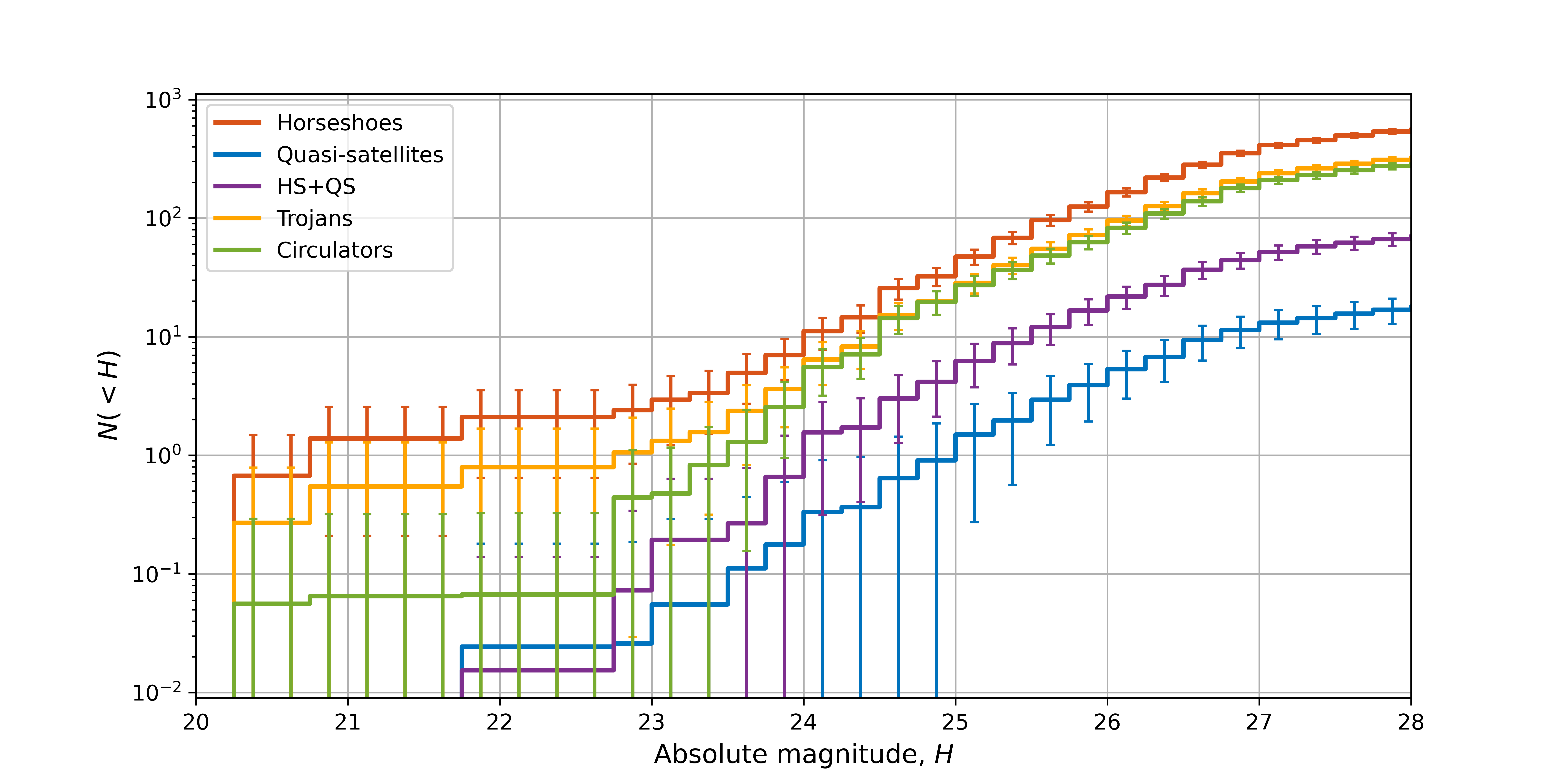}
    \caption{Estimated number of quasi-satellites (blue), horseshoes (red), HS+QS (violet), Trojans (orange), and circulators (green) as a function of the absolute magnitude $H$, obtained by combining the results presented here and the population from the NEOMOD3 model. }
    \label{fig:coorbital_numbers}
\end{figure}

Figure~\ref{fig:coorbital_numbers} shows the cumulative distribution of each co-orbital type, with bins of 0.25 width in absolute magnitude $H$. At $H=22$, which is the limit typically used to define potentially hazardous asteroids (PHAs), the total expected number of co-orbitals is about 3. At $H=25$, which corresponds to asteroids of $\sim$20-30 meters, about 78 co-orbitals predicted, of which $\sim$20 are Trojans, $\sim$40 are horseshoes, 3--4 HS+QS, and 1--2 quasi-satellites. Therefore, the Earth co-orbital population at this size is largely undiscovered. These numbers should be seen as order of magnitude estimates, since we did not estimate the average fraction of time spent in each co-orbital state, as done in \citep{fenucci-etal_2026}. However these estimates already highlight biases in the discovery of Earth co-orbitals. 

% Where are all the Trojans? 
Quasi-satellite orbits spend most of their time near $\sigma = 0$ deg and at opposition, making them ideal targets for discovery by ground-based surveys. Survey simulations performed by \citet{fenucci-etal_2026} showed that the ten-year operational lifetime of the Pan-STARRS survey should have detected roughly 80\% of Earth quasi-satellites up to absolute magnitude 24, while the upcoming Vera Rubin Observatory is expected to discover nearly the entire population up to magnitude 25. In contrast, discovering other type of co-orbital is generally more difficult \citep{morais-morbidelli_2002}, because these objects spend most of their time away from Earth and typically at low solar elongation. In addition, the discovery efficiency of the whole set of Earth co-orbitals is currently not well constrained in the literature. Systematic searches for Earth Trojans have been performed by using images from the Dark Energy Camera (DECAM) \citep{markwardt-etal_2020, lifset-etal_2021}, finding an upper limit of 938 Earth Trojans at $H=22$. A more recent study by \citep{yaeger-etal_2024} modeled the Earth Trojans population and the discovery efficiency of the Zwicky Transient Facility (ZTF), and found an upper limit of hundreds of objects up to $H=22$. Still, these number at larger than those coming from the NEA population modeling.
\citet{delafuente-delafuente_2015} investigated observational conditions of Earth co-orbitals at low eccentricity and low inclination, providing probability maps for their position in the plane of sky, providing constraints for survey pointings. The same work also argued that the low relative velocity with Earth at close encounters enlarges the impact cross section by gravitational focusing by a factor of 10 to 1000, thus generally increasing the impact probability of this subpopulation compared to the whole population of NEAs.

Expanding survey simulations to the whole population of Earth co-orbitals would provide valuable insights into the expected completeness of current and future surveys, especially to understand the expectations from the Vera Rubin telescope ten-years operational lifetime. 
In the next decade, infrared space-based surveys dedicated to the discovery of NEA, such as NEO Surveyor \citep{mainzer-etal_2023} by NASA and NEOMIR \citep{conversi-etal_2023} by ESA, are also foreseen. Because of their different observational technology and their geometry in the space, these missions are able to observe at lower solar elongations than ground-based telescopes, thus evaluating their contribution to the discovery of Earth co-orbitals would also be an asset.

\section{Conclusions}
\label{s:conclusions}
In this work, we have investigated the structure and short-term stability of the Earth co-orbital region at low eccentricity and low inclination by means of a semi-analytical resonant model and complementary full $N$-body numerical integrations. Using the resonant semi-secular Hamiltonian for the 1:1 mean-motion resonance, we characterized the topology of the phase space and quantified the fraction of different types of co-orbital motion.

Our analysis shows that the Earth co-orbital region is largely dominated by horseshoe orbits, which occupy more than half of the available phase space. Trojan orbits represent the second most frequent class, accounting for about one quarter of the volume. Quasi-satellite orbits are found to be intrinsically rare from a geometrical point of view, filling only a few percent of the total phase space volume. These results provide a quantitative description of the instantaneous resonant structure and clarify the relative geometric availability of different co-orbital states.
We also found that the phase space shows strong inhomogeneities, especially as a function of the argument of perihelion. Horseshoe and quasi-satellite orbits preferentially concentrate around $\omega = 90, \ 270$ deg, while circulators dominate near $\omega = 0, \ 180$ deg. These patterns are closely related to the node-crossing geometry with Earth. Trojan orbits, on the other hand, display an almost uniform distribution, consistent with the fact that conjunctions with Earth are dynamically forbidden for this class of motion.

The short-term stability analysis based on the MEGNO chaos indicator reveals that large portions of the Earth co-orbital region are dynamically chaotic on a timescale of 1000 years. High levels of chaos are primarily associated with close encounters with Earth and are particularly pronounced at low eccentricity and inclination, as well as near the node-crossing configurations. Among the different co-orbital classes, horseshoe and Trojan orbits appear to be, on average, the most chaotic, followed by circulators, while quasi-satellites show comparatively higher short-term stability, although they are not immune to chaotic evolution and state transitions.

From a planetary defence perspective, our results highlight that Earth co-orbitals cannot be regarded as dynamically protected from impact risk, and it is important to discover this particular population of NEAs. By combining our phase space fractions with a modern debiased NEA population model, we estimated the expected number of Earth co-orbitals as a function of size. We find that the currently known population represents only a small fraction of the expected one.
Overall, this study provides a global and quantitative picture of the resonant structure, stability properties, and population implications of Earth co-orbital asteroids.

\section*{Acknowledgments}
We thank the referees for their constructive comments which helped us improving the quality of the manuscript. The authors were also supported by the Spanish grant PID2024-158570NB-I00 (MICIU/\allowbreak AEI/\allowbreak10.13039/\allowbreak501100011033/\allowbreak FEDER/UE).
\bibliographystyle{apalike85}

\appendix

\section{Collision condition in the 1:1 mean-motion resonance}
\label{app:collision_angle}
Let us assume the hypotheses of Sec.~\ref{s:methods}, i.e. the planet is placed on a circular and planar orbit, and let us assume the asteroid is in an inclined orbit. Then, the collision can happen only at node crossing. The mean anomaly $\ell_p$ of the planet can be only $\ell_p = \Omega$ for a collision at the ascending node, or $\ell_p = \Omega - \pi$ for a collision at the descending node. 
In addition, the true anomaly $f$ of the asteroid can be only $f_{\rm coll} = -\omega$ for a collision at ascending node, of $f_{\rm coll} = \pi - \omega$ for a collision at the descending node.
Substituting these relations into $\sigma$, we get that the value corresponding to collision satisfies
\begin{equation}
    \sigma_{\text{c}} = \ell + \omega + \Omega - \ell_p = \ell - f_{\rm coll}.
    \label{eq:sigma_general}
\end{equation}
By the node crossing condition $\cos\omega=e$, we also have $\sin\omega = \sqrt{1-e^2}$ and 
\begin{equation}
    \tan\frac{\omega}{2} = \sqrt{\frac{1-e}{1+e}}.
    \label{eq:tanomega}
\end{equation}

\paragraph{Ascending Node}

At the ascending node, $f_{\rm coll} = -\omega$. The eccentric anomaly $E$ is obtained from
\begin{equation}
    \tan\frac{E}{2} = \sqrt{\frac{1-e}{1+e}}\tan\frac{f}{2}.
\end{equation}
Substituting $f = -\omega$ and using equation~\eqref{eq:tanomega}:
\begin{equation}
    \tan\frac{E}{2} = -\sqrt{\frac{1-e}{1+e}} \cdot \sqrt{\frac{1-e}{1+e}} = -\frac{1-e}{1+e}.
\end{equation}
Converting via the standard half-angle identities gives the exact expressions
\begin{equation}
    \cos E = \frac{2e}{1+e^2}, \qquad \sin E = -\frac{1-e^2}{1+e^2},
\end{equation}
and Kepler's equation then yields
\begin{equation}
    \ell = E - e\sin E = E + \frac{e(1-e^2)}{1+e^2}.
\end{equation}
The collision angle~\eqref{eq:sigma_general} is $\sigma_{\rm coll} = \ell + \omega$. Expanding all quantities as power series in $e$ near $e = 0$, using $E = -\pi/2 + 2e - \frac{4}{3}e^3 + \mathcal{O}(e^5)$ and $\omega = \pi/2 - e - \frac{1}{6}e^3 + \mathcal{O}(e^5)$, the leading $\pm\pi/2$ terms cancel exactly and one obtains
\begin{equation}
    \sigma_{\rm c}^{\rm asc} = 2e - \frac{7}{3}e^3 + \mathcal{O}(e^5).
    \label{eq:sigma_asc}
\end{equation}

\paragraph{Descending Node}

At the descending node, $f_{\rm coll} = \pi - \omega$. The half-angle formula gives
\begin{equation}
    \tan\frac{E}{2} = \sqrt{\frac{1-e}{1+e}}\tan\frac{\pi-\omega}{2} = \sqrt{\frac{1-e}{1+e}}\cot\frac{\omega}{2} = \sqrt{\frac{1-e}{1+e}} \cdot \sqrt{\frac{1+e}{1-e}} = 1,
\end{equation}
so that $E = \pi/2$ exactly, for all values of $e$. Consequently,
\begin{equation}
    \ell = \frac{\pi}{2} - e,
\end{equation}
and the collision angle is
\begin{equation}
    \sigma_{\rm c} = \ell - f_{\rm coll} = \left(\frac{\pi}{2} - e\right) - (\pi - \omega) = \omega - \frac{\pi}{2} - e.
\end{equation}
Expanding $\omega$ near $e = 0$ again cancels the $\pm\pi/2$ terms, yielding
\begin{equation}
    \sigma_{\rm c}^{\rm desc} = -2e - \frac{1}{6}e^3 + \mathcal{O}(e^5).
    \label{eq:sigma_desc}
\end{equation}

As a consequence, for circular orbits the collision can happen only at $\sigma_{\rm c}=0$ deg, thus circular quasi-satellite orbits do not exist. 

\bibliography{holybib}{}

\end{document}